\documentclass[journal]{IEEEtran}

\usepackage{amsmath}
\usepackage{amssymb}
\usepackage{amsfonts}
\usepackage{amsthm}
\usepackage{graphicx}
\usepackage{subfigure}
\usepackage{color}
\usepackage{eurosym}

\usepackage{graphicx}
\usepackage{pgfplots}
\pgfplotsset{compat=newest}
\usepackage{tikz}
\usetikzlibrary{patterns}

\theoremstyle{definition}

\newtheorem{remark}{Remark}

\usepackage{subfigure}

\usepackage{algorithm}
\usepackage[noadjust]{cite}

\usepackage{ctable}
\newcolumntype{M}[1]{>{\centering\arraybackslash}m{#1}}
\usepackage{colortbl}

\usepackage{multirow}
\usepackage{color}

\usepackage{algorithm}
\usepackage{algorithmic} 

\definecolor{airforceblue}{rgb}{0.36, 0.54, 0.66}

\newcommand{\sumg}{\sum\limits_{g \in \mathcal{G}}}

\newcommand{\sumd}{\sum\limits_{d \in \mathcal{D}}}
\newcommand{\sume}{\sum\limits_{e \in \mathcal{E}}}
\newcommand{\sumt}{\sum\limits_{t \in \mathcal{T}}}

\begin{document}

\title{Machine Learning for Improved Gas Network Models in Coordinated Energy Systems}

\author{Adriano~Arrigo,~\IEEEmembership{Student Member,~IEEE,}
Mihály~Dolányi,~\IEEEmembership{Student Member,~IEEE,}
Kenneth~Bruninx,~\IEEEmembership{Member,~IEEE}
and~Jean-François~Toubeau,~\IEEEmembership{Member,~IEEE}
%
}



\maketitle

\begin{abstract}
The current energy transition promotes the convergence of operation between the power and natural gas systems. In that direction, it becomes paramount to improve the modeling of non-convex natural gas flow dynamics within the coordinated power and gas dispatch. In this work, we propose a neural-network-constrained optimization method which includes a regression model of the Weymouth equation, based on supervised machine learning. The Weymouth equation links gas flow to inlet and outlet pressures for each pipeline via a quadratic equality, which is captured by a neural network. The latter is encoded via a tractable mixed-integer linear program into the set of constraints. In addition, our proposed framework is capable of considering bidirectionality without having recourse to complex and potentially inaccurate convexification approaches. We further enhance our model by introducing a reformulation of the activation function, which improves the computational efficiency. An extensive numerical study based on the real-life Belgian power and gas systems shows that the proposed methodology yields promising results in terms of accuracy and tractability.
\end{abstract}


%
\IEEEpeerreviewmaketitle


\section{Introduction}

\IEEEPARstart{T}{he} need to address climate change is triggering major transitions in the energy sector, one of which pertains to the increased integration of renewables in energy systems. In that context, Natural Gas-Fired Power Plants (NGFPPs) are becoming key flexible resources for cost-effectively managing the uncertainty and variability of weather-dependent renewable energy sources \cite{Hibbard2012Interdependence}. Moreover, the emergence of power-to-gas technologies that use electrical power to produce synthetic hydrogen or methane, paves the way towards the procurement of additional flexibility from gas storage, e.g., inside pipelines (which is referred to as linepack) or in gas tanks \cite{Chuan2017P2G}. In light of these inherent techno-economic dependencies between power and gas systems, there is an increasing need to improve the coordination of their operation \cite{Ameli20}. 
 


\color{black}
In this direction, several coordination schemes with different degrees of integration have been investigated in the literature \cite{Zlotnik2017Coordinated}. A first solution, which is regulatory-compliant as it does not introduce a new market mechanism, consists in making the day-ahead electricity market aware of gas network restrictions \cite{Byeon20}. A second potential solution, which increases the level of coordination between both energy systems, refers to the exchange of information (e.g., either based on volume or price quantities) between the electricity and gas market operators \cite{ORDOUDIS20201105}. Alternatively, the \textit{fully} coordinated day-ahead power and gas dispatch \cite{Shahidehpour1}, which is the focus of this paper, refers to a framework where a unique \textit{system operator} jointly dispatches the power and gas generating units. The resulting decision-making problem for system operator aims at minimizing the total day-ahead scheduling cost of both systems, while ensuring the demand- and supply-side operating constraints and network limitations. 






A major challenge which \textit{all} these works must contend with relates to the modeling of natural gas flow dynamics. In particular, the relationship between pressures and flow levels alongside the gas pipelines is described via a quadratic equality, the so-called Weymouth equation, which results in highly non-linear and non-convex constraints. In this work, to address this issue, we propose a neural-network-constrained coordinated day-ahead power and gas dispatch\footnote{In this work, we apply our proposed neural-network-constrained framework in a fully coordinated setting. It is worth emphasizing that this is not a limiting assumption, as our focus is on improving the modeling of gas flow dynamics and it is straightforward to incorporate our proposed framework within other coordination settings.}, which includes a regression model of the Weymouth equation (based on supervised machine learning) resulting in a tractable Mixed-Integer Linear Program (MILP). 

Different approaches exist in the literature to cope with the non-convexity of the Weymouth equation, which mainly rely on convexification techniques\footnote{Convexification techniques are of two main types \cite{CoffrinRoald}, namely, convex \textit{relaxations} (which enlarge the original feasible set), and convex \textit{approximations} (which may disregard some feasible solutions).}. Authors in \cite{AnubhavConicMarkets} use the Second-Order Cone (SOC) relaxation, which transforms the equality to an inequality constraint. However, this loose relaxation requires considering the whole interior of the quadratic cone instead of its envelope, which strongly jeopardizes the feasibility of the obtained solution. Hence, the SOC relaxation is supplemented with McCormick envelopes in \cite{Chen19}, which allow to exclude those solutions that yield a large feasibility gap with respect to the original Weymouth relation. To further improve the quality of the relaxation, authors in \cite{CoffrinStrengthening} propose tightening the bounds of the underlying McCormick envelopes, via an iterative algorithm which finds the smallest bounds around the operating point. However, this iterative procedure must be run online and results in a significant increase in computational burden. Generally speaking, all these SOC-based relaxation techniques are not sufficiently tight for practical purpose, and it is not straightforward to enhance these models for considering gas flow bidirectionality, see \cite{Shahidehpour2}. Differently, authors in \cite{BELDERBOS2020115082} and \cite{Correa2015Integrated} consider an interpolation of the Weymouth equation, via an incremental piecewise linear formulation, whose accuracy highly depends on the number of intervals selected to describe the non-linear function. The size of the resulting MILP increases with the number of intervals, therefore ensuing a trade-off between the approximation error and the computational burden \cite{Posada2014Comparison}. Overall, the previously explored convexification techniques still need to address fundamental research questions in terms of accuracy and tractability.




In order to reduce the inherent modeling errors arising from these convexification techniques, machine-learning-based optimization methods have been explored in the power system literature, with a special focus on Optimal Power Flow (OPF) problems\footnote{The OPF problems are exposed to non-convexities arising from, e.g., the modeling of AC power flows. The techniques (either based on convexification or machine learning approaches) to address these non-convexities are close to those required by the Weymouth equation, therefore making it important to review these works.}. In particular, authors in \cite{Canyasse2017} present a regression-based proxy of the solution to a real-time OPF, which is tailored to be embedded within long-term planning methodologies. Different regression tools are compared, shedding light on the appealing performances achieved by neural networks. Following the same rationale, authors in \cite{Xiang2021} and \cite{Fioretto2020} leverage a deep neural network architecture which calculates the optimal solution to a security-constrained DC-OPF and AC-OPF, respectively. These approaches reveal a substantial computational speed-up, however, their application in the scope of the power and gas dispatch problem is hindered by the number of operating points required in the training phase to yield an accurate model of the overall operating feasible space. To address this issue, several additional works consider the incorporation of machine learning models into the set of constraints. In particular, authors in \cite{Yeesian2018} aim at predicting the set of active constraints within the DC-OPF problem and propose a solution approach based on ensemble method to reduce the size of problem. Differently, the authors in \cite{SpyrosIEEENN} propose a \color{black} classification \color{black} neural-network-constrained AC-OPF encoding dynamic security restrictions. 
To the best of our knowledge, reference \cite{COSTA2016442} represents the first effort in the literature to implement a machine learning proxy within the power and gas coordinated energy systems\footnote{It is worth noting that decision trees have also been applied to OPF problems \cite{Cremer2019}-\hspace{1pt}\cite{Istemihan2010}.}. The authors leverage decisions trees as classifiers defining operating security regions. 




In this work, aiming at improving the modeling of non-convex gas flow dynamics, we propose a new neural-network-constrained optimization method. To that end, a neural network is trained offline in a fully data-driven way. In contrast to existing convex relaxation and approximation techniques, this training phase results in a \textit{regression} model linking the gas flow to inlet and outlet pressures for each pipeline, which minimizes the mean square error from the operating data points. Unlike machine learning models dedicated at directly finding the optimal decision set, our neural network is encoded via a MILP and embedded within the set of constraints, yielding a tractable, modular and comprehensible dispatch model. The proposed framework is further enhanced with a reformulation of activation function, which improves computational efficiency. 
%
%
Overall, the contributions of the work are as follows:
\begin{itemize}
  \item[\textit{(i)}] we formulate a neural-network-constrained coordinated day-ahead power and gas dispatch problem, including a regression model of the Weymouth equation, based on supervised machine learning. The resulting MILP provides a flexible framework enabling the consideration of bidirectional natural gas flow dynamics, without relying on any modeling assumption, 

  \item[\textit{(ii)}]  we improve the computational efficiency of the proposed approach by introducing a new reformulation of the activation function within the neural-network constrained power and gas dispatch problem. 
\end{itemize}

Our numerical analysis based on a case study inspired by the Belgian power and gas systems stresses that, the neural-network-constrained framework outperforms the existing SOC-based convexification techniques, and provides comparable performances with respect to piecewise linear approximation, for a similar number of mixed-integer variables involved in the optimization problem.





\section{The Coordinated Power and Gas Dispatch}\label{sec_gas_elec}
%
%
%
%
In the fully coordinated framework, the system operator aims at deriving a cost-efficient day-ahead scheduling of the electrical power and natural gas in-feeds. The dispatch solution must ensure operational restrictions pertaining to \textit{i)} the electrical network, \textit{ii)} the gas network, and \textit{iii)} the coupling of both, which are described in the following. 
%
\subsection{Electrical Network Constraints}
The electrical network connects conventional power generating units $e \in \mathcal{E}$, wind power generating units $w \in \mathcal{W}$, and electrical power demand centers $d \in \mathcal{D}$, via high voltage transmission lines $\ell \in \mathcal{L}$. Note that the set of power generating units also contains NGFPPs. The safe operation of electrical power network is ensured by the following set of constraints:
\begin{subequations}\label{ElecNet}
\begin{align}
& \hspace{-0.25cm} \sum_{e \in \mathcal{E}} p_{e,t} + \sum_{w \in \mathcal{W}} p_{w,t} = \sumd p_{d,t}, \enspace \forall t \in \mathcal{T}, \label{Elec1}\\
& \hspace{-0.25cm} \underline{P}_{e} \leqslant p_{e,t} \leqslant \overline{P}_{e}, \enspace \forall e \in \mathcal{E}, \; t \in \mathcal{T}, \label{Elec2}\\
& \hspace{-0.25cm} M_{\ell}^{\mathcal{E}} p_{e,t} + M_{\ell}^{\mathcal{W}} p_{w,t} - M_{\ell}^{\mathcal{D}} p_{d,t} \leqslant F^{\text{max}}_{\ell}, \enspace \forall {\ell} \in  \mathcal{L}, \; t \in \mathcal{T}. \label{Elec3} 
\end{align}
\end{subequations}

The constraint \eqref{Elec1} balances the total power generation $p_{e,t}$ and $p_{w,t}$ from the conventional and wind power generating units with the power demand $p_{d,t}$ from demand centers\footnote{In this work, we assume inelastic demand and deterministic wind power generation. It is worth mentioning that other works in the literature explore uncertainty-aware scheduling, such as, \cite{Qadrdan}, \cite{RATHA2020106565} and \cite{ARRIGO2022304}.}. The minimum and maximum operating limits $\underline{P}_e$ and $\overline{P}_e$ for each conventional power generating unit is enforced via \eqref{Elec2}. In \eqref{Elec3}, a DC load flow model \cite{Dvijotham2016DC} is leveraged for imposing the transmission line capacity limits $F^{\text{max}}_{\ell}$, using power transfer distribution factors, which map the contribution to the power flow in each line $\ell \in \mathcal{L}$ with the nodal injections. The latter are collected within $M_{\ell}^{\mathcal{E}}$, $M_{\ell}^{\mathcal{W}}$ and $M_{\ell}^{\mathcal{D}}$, for conventional power generating units $e \in \mathcal{E}$, wind power generating units $w \in \mathcal{W}$ and electrical power demand centers, respectively.


\subsection{Gas Network Constraints}

The gas network connects natural gas suppliers $g \in \mathcal{G}$ to natural gas demand centers $b \in \mathcal{B}$, via a network comprising of high pressure nodes $m \in \mathcal{M}$, high pressure gas pipelines $(m,n) \in \mathcal{Z}$ and compressors $(m,n) \in \mathcal{C}$, which are defined via their corresponding adjacent nodes $m$ and $n$. The safe operation of the natural gas network is ensured by the following set of constraints:
\begingroup
\begin{subequations}
\begin{align}
& \hspace{-0.15cm} \underline{P}_g \leqslant p_{g,t} \leqslant \overline{P}_g, \enspace \forall g \in \mathcal{G}, \; t \in \mathcal{T}, \label{Gas1} \\
& \hspace{-0.15cm} \underline{\text{PR}}_m \leqslant pr_{m,t} \leqslant \overline{\text{PR}}_m \enspace \forall m \in \mathcal{M}, \; t \in \mathcal{T}, \label{Gas2}\\
& \hspace{-0.15cm} pr_{n,t} \leqslant pr_{m,t} \leqslant \Gamma_{m,n} pr_{n,t}, \; \forall (m,n) \in  \mathcal{C}, \; t \in \mathcal{T}, \label{Gas3} \\
& \hspace{-0.15cm} q_{m,n,t}^2 = K_{m,n}^2 \left( pr_{m,t}^2 - pr_{n,t}^2 \right), \enspace \forall (m,n) \in \mathcal{Z}, \; t \in \mathcal{T}, \label{Weymouth} \\
& \hspace{-0.15cm} q_{m,n,t} = \frac{q_{m,n,t}^\text{in} + q_{m,n,t}^\text{out}}{2} \enspace \forall (m,n) \in \mathcal{Z}, \; t \in \mathcal{T}, \label{Gas5}\\
& \hspace{-0.15cm} h_{m,n,t} = S_{m,n} \frac{pr_{m,t} + pr_{n,t}}{2} \enspace \forall (m,n) \in \mathcal{Z}, \; t \in \mathcal{T}, \label{Gas6}\\
& \hspace{-0.15cm} h_{m,n,t} = H^0_{m,n} + q_{m,n,t}^\text{in} - q_{m,n,t}^\text{out} \enspace \forall (m,n) \in \mathcal{Z}, \enspace t = 1, \label{Gas7}\\
& \hspace{-0.15cm} h_{m,n,t} = h_{m,n,t-1} + q_{m,n,t}^\text{in} - q_{m,n,t}^\text{out} \, \forall (m,n) \in \mathcal{Z}, \; t \in \mathcal{T}_{\setminus \left\lbrace 1 \right\rbrace}. \label{Gas8} 
\end{align}
\end{subequations}

The constraint \eqref{Gas1} enforces the minimum and maximum operating limits $\underline{P}_g$ and $\overline{P}_g$ for gas supply $p_{g,t}$. The constraint \eqref{Gas2} enforces the nodal pressures $pr_{m,t}$ to lie within their minimum and maximum thresholds $\underline{\text{PR}}_m$ and $\overline{\text{PR}}_m$. In addition, the pressure at each adjacent nodes $(m,n) \in \mathcal{C}$ of each compressor is constrained by the constraint \eqref{Gas3}, involving the compression factor $\Gamma_{m,n}$. Equality constraint \eqref{Weymouth} corresponds to the Weymouth equation which provides the relation between the flow $q_{m,n,t}$ alongside the pipeline and the pressures $pr_{m,t}$ and $pr_{n,t}$ at each adjacent nodes $(m,n) \in \mathcal{Z}$ given the Weymouth constant $K_{m,u}$. This relation is derived from the partial differential equation underpinning the \textit{conservation of momentum} \cite{BELDERBOS2020115082}. Equation \eqref{Gas5} defines the physical flow as the average between inlet and outlet flows $q_{m,n,t}^\text{in}$ and $q_{m,n,t}^\text{out}$ for each pipeline $(m,n) \in \mathcal{Z}$. The linepack $h_{m,n,t}$, i.e., the inherent storage of gas inside each pipeline, is defined by the set of equations \eqref{Gas6} to \eqref{Gas8}, which are derived from the \textit{conservation of mass} principle. The spatial conservation of mass corresponds to equation \eqref{Gas6} which links the value of linepack to the pressures at inlet and outlet adjacent nodes, given the pipeline constant $S_{m,n}$. The mass conservation through time is enforced via equations \eqref{Gas7} and \eqref{Gas8}, where $H^0_{m,n}$ represents the initial condition of linepack. 


\subsection{Coupling Constraint}
The power and gas systems are coupled via the operation of NGFPPs, which consume natural gas for producing electrical power. This is revealed via the following relationship $\forall m \in \mathcal{M}, \; t \in \mathcal{T}$
\begingroup
\allowdisplaybreaks
\begin{equation}\label{GasElecCoupling}
\sum\limits_{g \in \mathcal{A}_m^\mathcal{G}} \hspace{-0.1cm} p_{g,t} - \hspace{-0.1cm} \sum\limits_{e \in \mathcal{A}_m^\mathcal{E}} \hspace{-0.1cm} \eta_e p_{e,t} - \hspace{-0.3cm} \sum\limits_{\underset{\in \mathcal{Z}}{  n:\left(m,n\right)}} \hspace{-0.1cm} \left( q_{m,n,t}^\text{in} - q_{m,n,t}^\text{out} \right) = \sum_{b \in \mathcal{B}} p_{b,t}, 
\end{equation}
\normalsize
\endgroup


\noindent where the sets $\mathcal{A}^\mathcal{G}_m$ and $\mathcal{A}_m^\mathcal{E}$ respectively collect the gas suppliers and the NGFPPs connected to node $m$, while the set $n:\left(m,n\right) \in \mathcal{Z}$ collects the pipelines connected to node $m$. Equation \eqref{GasElecCoupling} balances the gas supplies $p_{g,t}$ and inlet flows $q_{m,n,t}^\text{in}$ with the gas consumption by NGFPPs $\eta_e \, p_{e,t}$, outlet flows $q_{m,u,t}^\text{out}$, and gas demands $p_{b,t}$. The conversion factor $\eta_e$ defines the link between the energy consumed and that produced by the NGFPPs. 

\subsection{Resulting Model}
The resulting coordinated day-ahead power and gas dispatch problem writes as
\begin{subequations}\label{GasElecModel}
\begin{align}\label{GasElec}
& \min_{\Theta} \, \sumt \left( \sume C_{e} p_{e,t} +  \sumg C_{g} p_{g,t} \right),\\
& \text{s.t.} \enspace \text{Electrical Network Constraints \eqref{Elec1}-\eqref{Elec3}}, \\
& \hphantom{ \text{s.t.}} \enspace \text{Gas Network Constraints \eqref{Gas1}-\eqref{Gas8}}, \\
& \hphantom{ \text{s.t.}} \enspace \text{Coupling Constraint \eqref{GasElecCoupling}},
\end{align}
\end{subequations}
\noindent where $\Theta = \left\lbrace p_{e,t}, \, p_{g,t}, \, pr_{m,t}, \, q_{m,u,t}, \, q^{\text{in}}_{m,u,t}, \, q^{\text{out}}_{m,u,t}, \, h_{m,u,t} \right\rbrace$ represent the set of decision variables. By solving problem \eqref{GasElec}, the system operator aims at minimizing the total day-ahead cost of scheduling the electrical power and natural gas in-feeds, where $C_e$ and $C_g$ respectively represent the marginal generation cost of electricity and natural gas, while ensuring the operating constraints of both power and gas systems.

\section{The Neural-Network-Constrained Framework}\label{sec_methodo}


The Weymouth equation \eqref{Weymouth} is a non-convex quadratic equality constraint, which highly hinders the resolution of the day-ahead coordinated power and gas dispatch problem \eqref{GasElecModel} via off-the-shelf solvers. In this work, we propose a neural-network-constrained optimization method which allows to recover tractability, by encoding a regression model of the Weymouth equation as a MILP into the set of constraints. In the following, we explain the methodology which requires \textit{i)} to train the neural network offline (Section \ref{sec_NN_train}), and \textit{ii)} to formulate the regression model as a MILP (Section \ref{subsect_refor_LR}). Section \ref{subsect_NNPGmodel} introduces the resulting neural-network-constrained coordinated power and gas dispatch problem to solve. 



    

\subsection{Training the Neural Network}\label{sec_NN_train}




\begin{figure}
\centering
\includegraphics[width=0.95\columnwidth]{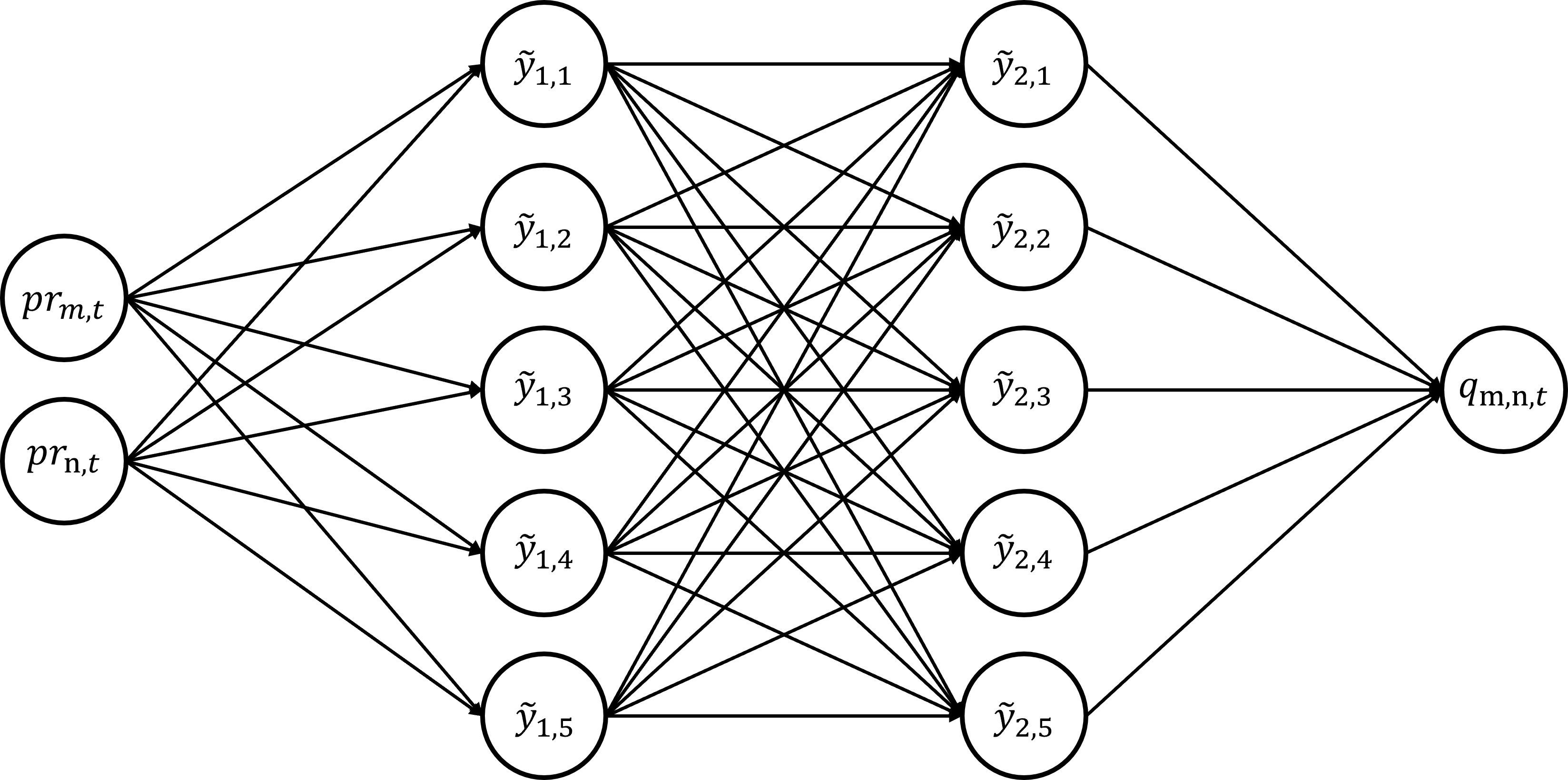}
\caption{An illustrative neural network for the Weymouth equation, linking the inlet and outlet pressures to the flow. The neural network is composed of two hidden layers and five neurons per layer. Each branch is assigned with a weight which defines the weighted sum, fed to the activation function.}
\label{fig:NeuralNetwork}
\end{figure}


Neural networks represent a type of supervised machine learning regression tool that is capable of extracting non-linear patterns from data, that would not be captured by other regression techniques (e.g., linear or polynomial regression). Their architecture is composed of neurons that are interconnected between each other via weighting branches. An example of such architecture, composed of 2 layers and 5 neurons per layer, is represented in Figure \ref{fig:NeuralNetwork} for the specific case of Weymouth equation, which relates the inlet and outlet pressures with the gas flow within the pipeline. The fundamental building block of the neural network is the neuron, i.e., a non-linear activation function $g : \mathbb{R} \rightarrow \mathbb{R}$ whose input is the weighted sum of outputs from the previous layer, and a bias term. There exist different types of activation functions that may be leveraged, such as the sigmoid, the hyperbolic tangent, the ReLU, and the leaky ReLU which are shown in Figure \ref{fig:Activation functions}. 
%


The operation of the neural network can be written in a mathematical vectorial form, such as
\begin{subequations}\label{NNOperatingEquations}
\begin{align}
& \widetilde{y}_{k+1} = W_{k+1} \, y_k + B_{k+1} \enspace \forall k \in \left\lbrace 0, 1, ..., K-1 \right\rbrace, \label{NNOperatingEquations1}\\
& y_{k} = g \left( \widetilde{y}_{k} \right) \enspace \forall k \in \left\lbrace 1, ..., K \right\rbrace, \label{NNOperatingEquations2}\\
& z = W_{K+1} \, y_K + B_{K+1}. \label{NNOperatingEquations3}
\end{align}
\end{subequations}

In \eqref{NNOperatingEquations}, the input variables are $y_0 = \left[ pr_{m,t} \enspace pr_{n,t} \right]^\top$ and each layer $k \in \left\lbrace 1, ..., K \right\rbrace$ is composed of $N_k$ neurons. The weighted sum $\widetilde{y}_k \in \mathbb{R}^{N_k}$ is calculated in \eqref{NNOperatingEquations1} via the weight matrices $W_k \in \mathbb{R}^{N_k \times N_{k-1}}$ and the bias vector $B_k \in \mathbb{R}^{N_k}$. It is then fed to the activation function $g \left( \widetilde{y}_k \right)$ in equation \eqref{NNOperatingEquations2}. \color{black} Note that we define $g(.): \mathbb{R}^{N_k} \rightarrow \mathbb{R}^{N_k}$ as a vector function, whose input and output are vectors of the same length, the elements of which are linked via the underlying unidimensional activation function. \color{black} Finally, the output of the network $z = q_{m,n,t}$ is calculated by \eqref{NNOperatingEquations3}, where $W_{K+1} \in \mathbb{R}^{1 \times N_k}$ and $B_{K+1} \in \mathbb{R}$, respectively represent the weight vector and the bias term of the output layer. We compactly write the resulting non-linear regression function as $z = f \left( y_0, W, B \right) $, where $y_0 $ is the input vector, and $W, B$ respectively collect the weight matrices $W_k$ and the bias vectors $B_k$, for each layer $k$ and the output layer.

The challenge for utilizing the neural network \eqref{NNOperatingEquations} is to appropriately select the weights $W$ and biases $B$, during the so-called \textit{training} phase. Assuming the availability of a training dataset $\left\lbrace \widehat{y}_{0,i}, \widehat{z}_i \right\rbrace \enspace \forall i \in \left\lbrace 1, ..., N \right\rbrace$,  this task is usually achieved by solving the following optimization problem 
\begin{align}\label{LossMinimization}
\min_{W, B} \enspace \frac{1}{N} \sum\limits_{i = 1}^N \mathcal{L} \left( f \left( \widehat{y}_{0,i}, W, B \right) , \widehat{z}_i \right).
\end{align}

\begin{figure}\small
\begin{minipage}[]{0.49\columnwidth}
\centering
\begin{tikzpicture}[declare function={f(\x)= 1/(1+exp(-\x)); }, declare function={g(\x)= (exp(\x) - exp(-\x))/(exp(\x)+exp(-\x));},
declare function={h(\x)= (\x<0) * 0 + (\x>0) * x;} ]
\pgfplotsset{width=\columnwidth}
\begin{axis}[
		grid = major,
    		xtick={-5,0,5},
    		ytick={-1,0,1},
  		axis lines=left,
  		xlabel={$\widetilde{y}$}, ylabel={$g \left( \widetilde{y} \right)$},
  		domain=-5:5,
		xmin=-5, xmax=5,
  		ymin=0, ymax=1, 
  		samples=20,
 ]
  \addplot[mark=none, line width=1.5, color=airforceblue]
  	{f(x)};
\end{axis}
\end{tikzpicture}
\end{minipage}
\begin{minipage}[]{0.49\columnwidth}
\centering
\begin{tikzpicture}[declare function={f(\x)= 1/(1+exp(-\x)); }, declare function={g(\x)= (exp(\x) - exp(-\x))/(exp(\x)+exp(-\x));},
declare function={h(\x)= (\x<0) * 0 + (\x>0) * x;} ]
\pgfplotsset{width=\columnwidth}
\begin{axis}[
		grid = major,
    		xtick={-5,0,5},
    		ytick={-1,0,1},
  		axis lines=left,
  		xlabel={$\widetilde{y}$}, ylabel={$g \left( \widetilde{y} \right)$},
  		domain=-5:5,
		xmin=-5, xmax=5,
  		ymin=-1, ymax=1, 
  		samples=20,
 ]
  \addplot[mark=none, line width=1.5, color=airforceblue]
  	{g(x)};
\end{axis}
\end{tikzpicture}
\end{minipage}
\begin{minipage}[]{0.49\columnwidth}
\centering
\begin{tikzpicture}[declare function={f(\x)= 1/(1+exp(-\x)); }, declare function={g(\x)= (exp(\x) - exp(-\x))/(exp(\x)+exp(-\x));},
declare function={h(\x)= (\x<0) * 0 + (\x>0) * x;} ]
\pgfplotsset{width=\columnwidth}
\begin{axis}[
		grid = major,
    		xtick={-5,0,5},
    		ytick={-5,0,1,2,3,4,5},
  		axis lines=left,
  		xlabel={$\widetilde{y}$}, ylabel={$g \left( \widetilde{y} \right)$},
  		domain=-5:5,
		xmin=-5, xmax=5,
  		ymin=0, ymax=5, 
  		samples=20,
 ]
  \addplot[mark=none, line width=1.5, color=airforceblue]
  	{h(x)};
\end{axis}
\end{tikzpicture}
\end{minipage}
\begin{minipage}[]{0.49\columnwidth}
\centering
\begin{tikzpicture}[declare function={f(\x)= 1/(1+exp(-\x)); }, declare function={g(\x)= (exp(\x) - exp(-\x))/(exp(\x)+exp(-\x));},
declare function={h(\x)= (\x<0) * 0.2*x + (\x>0) * x;} ]
\pgfplotsset{width=\columnwidth}
\begin{axis}[
		grid = major,
    		xtick={-5,0,5},
    		ytick={-4,-2,0,2,4},
  		axis lines=left,
  		xlabel={$\widetilde{y}$}, ylabel={$g \left( \widetilde{y} \right)$},
  		domain=-5:5,
		xmin=-5, xmax=5,
  		ymin=-2, ymax=5, 
  		samples=20,
 ]
  \addplot[mark=none, line width=1.5, color=airforceblue]
  	{h(x)};
\end{axis}
\end{tikzpicture}
\end{minipage}
\caption{From top left to bottom right: the sigmoid, hyperbolic tangent and ReLU and leaky ReLU activation functions.}
\label{fig:Activation functions}
\end{figure}
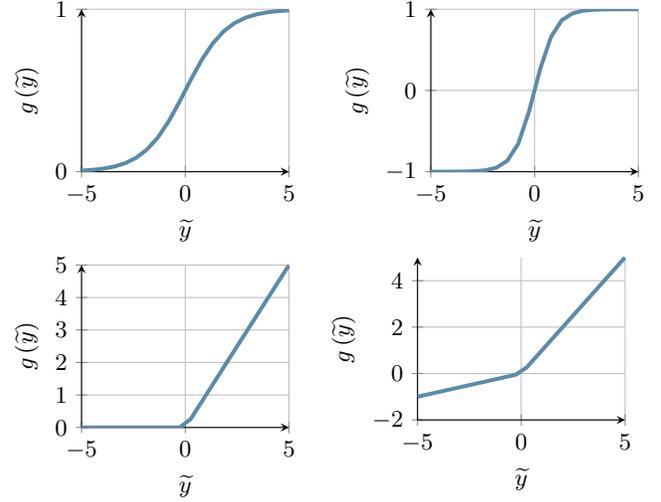
\normalsize

Problem \eqref{LossMinimization} aims at minimizing the expected value of loss function over the training dataset. The loss function $\mathcal{L} \left( . , . \right) $ gives a score for the error between the neural network output $f \left( \widehat{y}_{0,i}, W, B \right) $ and the expected one $ \widehat{z}_i $. In this work, we consider the square error as our loss function, as it is the common practice for regression models in the literature. Problem \eqref{LossMinimization} is non-linear, due to the loss function but also due to the neural network transfer function. For solving \eqref{LossMinimization}, we use dedicated stochastic gradient descent algorithm and backpropagation techniques, that are well documented in the literature \cite{Goodfellow2016}.

The set of equations \eqref{NNOperatingEquations} equipped with the optimal weight matrices obtained from \eqref{LossMinimization}, is a regression model for the Weymouth equation. In the following, our aim is to incorporate \eqref{NNOperatingEquations} into the set of constraints of the coordinated power and gas dispatch problem, to replace the non-convex Weymouth equation. The activation function in \eqref{NNOperatingEquations2} therefore requires reformulation. In that direction, we propose a tractable MILP reformulation of the activation function in the next Section \ref{subsect_refor_LR}.

\subsection{Reformulation of the Activation Function}\label{subsect_refor_LR}
The use of non-linear activation functions is essential for enabling neural networks to learn complex mapping functions. In this way, the adoption of the Rectifier Linear Unit (ReLU) and leaky ReLU is one of the key milestones in the advent of machine learning \cite{Glorot11a}, as it has been shown that neural networks relying on these functions are easier to train\footnote{The ReLU and leaky ReLU activation functions are nearly linear, and are therefore easy to optimize within gradient-based techniques.} and often outperform traditional sigmoid and hyperbolic tangent activation functions \cite{LeCun15}. Following these strong empirical evidences, we propose a tractable MILP reformulation of these activation functions within the proposed neural-network-constrained framework. We begin with the definition of ReLU and leaky ReLU. Mathematically speaking, these functions may be written in the form of 
\begin{align}
    & \text{ReLU} \left( \widetilde{y} \right) = \max \left( 0, \widetilde{y} \right), \label{ReLU} \\
    & \ell\text{-ReLU} \left( \widetilde{y} \right) = \max \left( \alpha \widetilde{y} , \widetilde{y} \right), \label{l-ReLU}
\end{align}

\noindent where $\text{ReLU} \left( . \right) : \mathbb{R} \rightarrow \mathbb{R} $ and $\ell\text{-ReLU} \left( . \right) : \mathbb{R} \rightarrow \mathbb{R} $ respectively represent the ReLU and leaky ReLU activation functions. The parameter $\alpha \in \left[ 0, 1 \right]$ defines the slope of the left-hand side non-positive part of the leaky ReLU activation function. Note that selecting $\alpha = 0$ for $\ell\text{-ReLU}$ amounts to deriving the ReLU activation function. Therefore, we propose the following MILP reformulation as a generic model for encoding \eqref{NNOperatingEquations2}, and for both functions:
%
%
\begin{subequations}\label{ReforLeakyRelu}
\begin{align}
    & \widetilde{y}_k^\text{min} b_k \leqslant x_{1,k} \leqslant 0, & \forall k \in \left\lbrace 1, ..., K \right\rbrace, \label{LR1}\\
    & 0 \leqslant x_{2,k} \leqslant \widetilde{y}_k^\text{max} ( 1 - b_k ), & \forall k \in \left\lbrace 1, ..., K \right\rbrace, \label{LR2}\\
    & \widetilde{y}_k = x_{1,k} + x_{2,k}, & \forall k \in \left\lbrace 1, ..., K \right\rbrace, \label{LR3}\\
    & y_k = \alpha x_{1,k} + x_{2,k}, & \forall k \in \left\lbrace 1, ..., K \right\rbrace, \label{LR4}\\
    & b_k \in \left\lbrace 0, 1 \right\rbrace^{N_k}, & \forall k \in \left\lbrace 1, ..., K \right\rbrace, \label{LR5}
    \end{align}
\end{subequations}

\noindent where all the neurons in \eqref{NNOperatingEquations2} are simultaneously considered, and $x_{1,k} \in \mathbb{R}^{N_k}$, $x_{2,k} \in \mathbb{R}^{N_k}$ and $b_{k} \in \left\lbrace 0, 1 \right\rbrace^{N_k}$ are auxiliary continuous and binary variables. Note that we use vector notation, for the sake of clarity. The MILP \eqref{ReforLeakyRelu} works as follows. The binary variables $b_k$ equal to 0 or 1, respectively when the negative or positive part of the function is active. Therefore, the combination of equations \eqref{LR1}, \eqref{LR2} and \eqref{LR3}, implies that $x_{1,k}$ and $x_{2,k}$ respectively take the negative and positive value of $\widetilde{y}_k$. The parameters $\widetilde{y}_k^\text{min} \in \mathbb{R}^{N_k}$ and $\widetilde{y}_k^\text{max} \in \mathbb{R}^{N_k}$, collect the minimum and maximum admissible values for the input $\widetilde{y}_k$ to each neuron. The equation \eqref{LR4} calculates the output of the activation function, given the values of $x_{1,k}$ and $x_{2,k}$. Finally, equation \eqref{LR5} defines $b_k$ as a vector of binary variables. In Section \ref{subsect_NNPGmodel}, we incorporate the MILP \eqref{ReforLeakyRelu} into the set of constraints of the power and gas dispatch problem to improve the modeling of gas flow dynamics.

\begin{remark}[Benchmarking the proposed reformulation of activation function] To appropriately show the superiority of the proposed reformulation of activation function, we compare the approach with convexification techniques (which will be presented later in Section \ref{sec_base}) as well as with an existing reformulation of ReLU proposed in \cite{SpyrosIEEENN}, which is as follows:
%
%
\begin{subequations}\label{ReforRelu}
\begin{align}
    & \widetilde{y}_k \leqslant y_k \leqslant \widetilde{y}_k - \widetilde{y}_k^\text{min} ( 1 - b_k ), & \forall k \in \left\lbrace 1, ..., K \right\rbrace, \label{R1}\\
    & 0 \leqslant y_k \leqslant \widetilde{y}_k^\text{max}  b_k, & \forall k \in \left\lbrace 1, ..., K \right\rbrace, \label{R2}\\
    & b_k \in \left\lbrace 0, 1 \right\rbrace^{N_k}, & \forall k \in \left\lbrace 1, ..., K \right\rbrace. \label{R3}
    \end{align}
\end{subequations}

In \eqref{ReforRelu}, the binary variables $b_k \in \left\lbrace 0, 1 \right\rbrace^{N_k}$ either activate or deactivate the right-hand side inequalities in \eqref{R1} and \eqref{R2}. By doing so, $b_k = 1$ implies $y_k = \widetilde{y}_k$, and $b_k = 0$ implies $y_k = 0$. It is worth mentioning that our reformulation in \eqref{ReforLeakyRelu} generalizes the approach in \cite{SpyrosIEEENN}, since it is capable of considering both ReLU and leaky ReLU. In addition, we will show later in our numerical experiments in Section \ref{sect:NumericalExperiments} that our reformulation outperforms \eqref{ReforRelu} in terms of computational efficiency.

\end{remark}

\subsection{The Neural-Network-Constrained Coordinated Power and Gas Dispatch}\label{subsect_NNPGmodel}

The reformulation proposed in Section \ref{subsect_refor_LR} for incorporating the neural-network constraints in the coordinated power and gas dispatch problem results in the following MILP 
\begin{subequations}\label{GasElecModel}
\begin{align}
& \min_{\Theta} \, \sumt \left( \sume C_{e} p_{e,t} +  \sumg C_{g} p_{g,t} \right),\\
& \text{s.t.} \enspace \text{Electrical Network Constraints \eqref{Elec1}-\eqref{Elec3}}, \\
& \hphantom{ \text{s.t.}} \enspace \text{Gas Network Constraints \eqref{Gas1}-\eqref{Gas3}, \eqref{Gas5}-\eqref{Gas8}}, \\
& \hphantom{ \text{s.t.}} \enspace \text{Coupling Constraint \eqref{GasElecCoupling}},\\
& \hphantom{ \text{s.t.}} \enspace \text{Neural network constraints \eqref{NNOperatingEquations1}, \eqref{NNOperatingEquations3}, and \eqref{ReforLeakyRelu},} \label{NNConstraint}
\end{align}
\end{subequations}

\noindent where the intractable non-convex Weymouth equation \eqref{Weymouth} is replaced by the tractable MILP \eqref{NNConstraint} encoding the neural network, via \eqref{NNOperatingEquations1}, \eqref{NNOperatingEquations3}, and the reformulation of activation function in \eqref{ReforLeakyRelu}.

\begin{remark}[Benchmarking the proposed reformulation of activation function]
The procedure to derive the neural-network-constrained power and gas dispatch problem, given the existing reformulation of ReLU activation function \cite{SpyrosIEEENN}, is similar to that explained in this section, except that \eqref{ReforRelu} is used instead of \eqref{ReforLeakyRelu}, to reformulate the activation function.
\end{remark}

\section{Benchmark Convexification Techniques}\label{sec_base}





In this section, we introduce existing convexification techniques for the non-convex Weymouth equation \eqref{Weymouth}, that will be used later as benchmark solutions in the numerical experiments of Section \ref{sect:NumericalExperiments}. In particular, we focus on the SOC relaxation in Section \ref{subsect:SOC}. Next, tightened versions of the SOC relaxation are presented in Sections \ref{subsect:SOCCorm} and \ref{subsect:SOCTigh}, respectively using McComick envelopes and a dedicated bound tightening algorithm. Finally, an incremental piecewise linear approximation is showcased in Section \ref{subsect:PWL}.

\subsection{SOC Relaxation}\label{subsect:SOC}

The most commonly used technique to address the non-convexity is to equivalenty cast the Weymouth equation into
\begin{align}
& q_{m,n,t}^2 \leqslant K_{m,n}^2 \left( pr_{m,t}^2 - pr_{n,t}^2 \right), \enspace \forall (m,n) \in \mathcal{Z}, \; t \in \mathcal{T}, \label{SOCWeymouth} \\
& q_{m,n,t}^2 \geqslant K_{m,n}^2 \left( pr_{m,t}^2 - pr_{n,t}^2 \right), \enspace \forall (m,n) \in \mathcal{Z}, \; t \in \mathcal{T}, \label{SOCWeymouthrelax} 
\end{align}

\noindent and relax the non-convex constraint \eqref{SOCWeymouthrelax}, thereby replacing the equality constraint by the SOC constraint in \eqref{SOCWeymouth}. The difference between the original Weymouth equation and the SOC constraint \eqref{SOCWeymouth} is illustrated in Figure \ref{fig:WeymouthSOC}. The left-hand side Fig. \ref{Fig:3DWeymouth} represents the original bidirectional link between natural gas flow and inlet and outlet pressures, captured by the non-convex conic surface. The right-hand side Fig. \ref{fig:SOCPRelaxation} represents the feasible space considered under the SOC relaxation (i.e., the whole grey volume under the curve). These graphs suggest that the solutions obtained from the SOC relaxation may bring non-trivial operational constraint violations. Therefore, additional tightening of the relaxation have been explored in the literature.

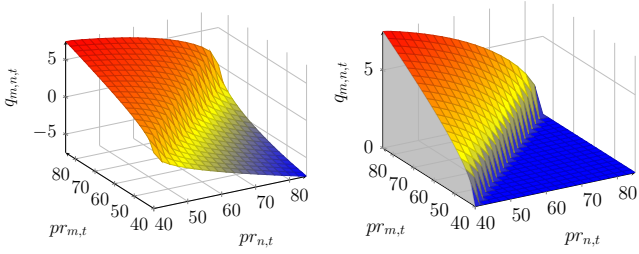
\begin{figure}
\centering
\Large
\subfigure[The Weymouth relation between inlet, outlet pressures $pr_{m,t}$, $pr_{n,t}$, and the flows $q_{m,n,t}$ within the pipeline. Pressures are expressed in bar, flows are expressed in MNm$^3$/h. \label{Fig:3DWeymouth}]{\resizebox {0.48\columnwidth} {!} {
\begin{tikzpicture}[declare function={ f(\x,\y)=
											(\x>\y) * 0.1*abs(\x^2-\y^2)^0.5
											+
											(\x<\y) * -0.1*abs(\y^2-\x^2)^0.5; }]
\begin{axis}[
		grid = major,
		view={60}{30},
    	xtick={40, 50, 60, 70, 80},
    	ytick={40, 50, 60, 70, 80},
  		axis lines=left,
  		xlabel={$pr_{m,t}$}, ylabel={$pr_{n,t}$}, zlabel={$q_{m,n,t}$},
  		domain=40:85,
  		y domain=40:85,
  		x dir=reverse,
		xmin=40, xmax=85,
  		ymin=40, ymax=85, zmin=-7.5, zmax=7.5,
  		samples=20,
  		samples y=20,
  		z buffer=sort,
  		height=7cm,
 			]
\addplot3[surf] {f(x,y)};
\end{axis}
\end{tikzpicture}}}
\subfigure[SOC relaxation. The grey area shows the area under the surface which is added to the feasible space by conic relaxation. The negative flows are neglected.\label{fig:SOCPRelaxation}]{\resizebox {0.48\columnwidth} {!} {
\begin{tikzpicture}[declare function={f(\x,\y)=(\x>\y) * 0.1*abs(\x^2-\y^2)^0.5;
  }]
\pgfdeclarelayer{pre main}
\pgfsetlayers{pre main,main}
\begin{axis}[
view={60}{30},
		grid = major,
		view={60}{30},
    		xtick={40, 50, 60, 70, 80},
    		ytick={40, 50, 60, 70, 80},
  		axis lines=left,
  		xlabel={$pr_{m,t}$}, ylabel={$pr_{n,t}$}, zlabel={$q_{m,n,t}$},
  		domain=40:85,
  		y domain=40:85,
  		x dir=reverse,
		xmin=40, xmax=85,
  		ymin=40, ymax=85, zmin=0, zmax=7.5,
  		samples=20,
  		samples y=20,
  		z buffer=sort,
  		height=7cm,
 ]
  \addplot3[surf]
  	{f(x,y)};
  \fill[color=white!70!black, opacity=0.8] (40,40,0) -- plot[variable=\x,domain=40:85] (\x,40,{f(\x,40)}) -- (85,40,0) -- cycle;
\end{axis}
\end{tikzpicture}}}
\caption{Three-dimensional plot of the Weymouth relation and its SOC relaxation.}
\label{fig:WeymouthSOC}
\end{figure}

\subsection{SOC Relaxation with McCormick Envelopes}\label{subsect:SOCCorm}

The first method used to tighten the SOC relaxation consists in introducing a convex relaxation of \eqref{SOCWeymouthrelax} using the McCormick envelopes, as proposed in \cite{Chen19}. By doing so, the volume under the curve which contains infeasible operating points is shrunk. Mathematically speaking, this amounts to replace \eqref{SOCWeymouthrelax} with the following set of equations, from which we omit ``$\forall \left( m,n \right) \in \mathcal{Z}, \; t \in \mathcal{T}$'' in each constraint, for the sake of conciseness:
\begin{subequations}\label{SOCPMcCormick}
\begin{align}
& \hspace{-0.2cm} \kappa_{m,n,t} \geqslant K_{m,n}^2 \lambda_{m,n,t} ,\\
& \hspace{-0.2cm} \kappa_{m,n,t} \geqslant q_{m,n,t}^2 ,\\
& \hspace{-0.2cm} \kappa_{m,n,t} \leqslant \left( Q_{m,n,t}^{\text{max}} + Q_{m,n,t}^{\text{max}} \right) q_{m,n,t} - Q_{m,n,t}^{\text{max}} Q_{m,n,t}^{\text{min}} ,\\
& \hspace{-0.2cm} \lambda_{m,n,t} \hspace{-0.05cm} \geqslant \hspace{-0.05cm} A^\text{min}_{m,n,t} b_{m,n,t} + B_{m,n,t}^\text{min} a_{m,n,t}- A_{m,n,t}^\text{min} B_{m,n,t}^\text{min} ,\\
& \hspace{-0.2cm} \lambda_{m,n,t} \hspace{-0.05cm} \geqslant \hspace{-0.05cm} A^\text{max}_{m,n,t} b_{m,n,t} + B_{m,n,t}^\text{max} a_{m,n,t} - A_{m,n,t}^\text{max} B_{m,n,t}^\text{max} ,\\
& \hspace{-0.2cm} \lambda_{m,n,t} \hspace{-0.05cm} \leqslant \hspace{-0.05cm} A^\text{min}_{m,n,t} b_{m,n,t} + B_{m,n,t}^\text{max} a_{m,n,t} - A_{m,n,t}^\text{min} B_{m,n,t}^\text{max} ,\\
& \hspace{-0.2cm} \lambda_{m,n,t} \hspace{-0.05cm} \leqslant \hspace{-0.05cm} A^\text{max}_{m,n,t} b_{m,n,t} + B_{m,n,t}^\text{min} a_{m,n,t} - A_{m,n,t}^\text{max} B_{m,n,t}^\text{min} .
\end{align}
\end{subequations}

In \eqref{SOCPMcCormick}, $\kappa_{m,n,t}$, $\lambda_{m,n,t}$, $a_{m,n,t}$ and $b_{m,n,t}$ are auxiliary variables. In addition, $a_{m,n,t} = pr_{m,t} + pr_{n,t}$ and $b_{m,n,t} = pr_{m,t} - pr_{n,t}$. In general, the convex relaxation \eqref{SOCPMcCormick} provides a more accurate representation of the feasible set than the one achieved by the SOC relaxation. However, it requires the introduction of the bounds $Q_{m,n,t}^\text{max}$, $Q_{m,n,t}^\text{min}$, $A_{m,n,t}^\text{max}$, $A_{m,n,t}^\text{min}$, $B_{m,n,t}^\text{max}$, and $B_{m,n,t}^\text{min}$, which are not straightforward to calculate. The introduction of loose bounds may still provide an insufficient representation of the Weymouth constraint.

\subsection{Tightening the Bounds of McCormick Envelopes}\label{subsect:SOCTigh}

To address this issue, the authors in \cite{Chen19}-\hspace{1pt}\cite{CoffrinStrengthening} propose a bound tightening algorithm that enables to improve the accuracy of the relaxation. The iterative algorithm is able to calculate effective bounds, i.e., that reduce the available set of solutions around the operating point. Once these bounds are derived, the power and gas dispatch problem with SOC relaxation and McCormick envelopes, can be solved efficiently. However, it is worth mentioning that the procedure to derive the bounds must be run online (i.e., different bounds are required under different operating conditions), therefore resulting in a significant increase in computational burden. For the sake of brevity, we present the flow of this algorithm in our online companion \cite{OnlineAppendix}.

\color{black}
\begin{remark}
In general, the SOC-based convex relaxation techniques can not straightforwardly enhance to consider bidirectionality. The previous tentatives \cite{SchweleCoordination} towards SOC-based relaxations of bidirectional gas flows usually rely on binary variables defining the direction of flow and big-M parameters. These modeling approaches may arise in infeasible and intractable dispatch solutions due to the impreciseness introduced by big-M parameters and the difficulty for branch-and-bound algorithms to quickly determine the direction of gas flow in all pipelines of the system.
\end{remark}
\color{black}

\subsection{Incremental Piecewise Linear Approximation}\label{subsect:PWL}


Piecewise linear functions represent an alternative potential approximation for the Weymouth equation. In particular, the incremental method, as described in \cite{BELDERBOS2020115082}-\hspace{1pt}\cite{Correa2015Integrated} has shown to outperform other existing piecewise linear models for modeling gas network \cite{Posada2014Comparison}. The main idea behind the incremental method is to linearize the non-linear terms that compose the Weymouth equation, i.e., square pressures $pr_{m,t}^2$ and square gas flows $q_{m,n,t}^2$. To do so, a collection of breakpoints are collected, i.e., $\left\lbrace Q_{j},  Q_{j}^{2} \right\rbrace$ and $\left\lbrace PR_{j}^{2}, PR_{j}^{2} \right\rbrace$ where $j \in \left\lbrace 1, ..., J \right\rbrace$, and auxiliary \vspace{0.1cm} variables $h^{q}_{m,n,t}$ and $h^{pr}_{m,t}$ are introduced to \vspace{0.1cm} describe the non-linear terms $q_{m,n,t}^2$ and $pr_{m,t}^2$, respectively. For instance, the following set of equation produces a piecewise linear approximation $h^{q}_{m,n,t}$ of $q_{m,n,t}^2$ from which we omit ``$\forall \left( m,n \right) \in \mathcal{Z}, \; t \in \mathcal{T}$'' in each constraint for the sake of conciseness:
%
\begin{subequations}\label{hqequation}
\begin{align}
& h^{q}_{m,n,t} = Q_{1}^{2} + \sum\limits_{j = 1}^{J} \left( Q_{j+1}^2 - Q_{j}^2 \right) \delta^q_{m,n,t,j}, & \label{hq1} \\
& q_{m,n,t} = Q_{1} + \sum\limits_{j = 1}^{J - 1} \left( Q_{j+1} - Q_{j} \right) \delta^q_{m,n,t,j}, &  \label{hq2} \\
& 0 \leqslant \delta^q_{m,n,t,j} \leqslant 1, &  \hspace{-3cm} \forall j \in \left\lbrace 1, ..., J \right\rbrace, \label{hq3}\\
& \varphi^q_{m,n,t,j} \leqslant \delta^q_{m,n,t,j},  & \hspace{-3cm} \forall j \in \left\lbrace 1, ..., J \right\rbrace, \label{hq4}\\
& \delta^q_{m,n,t,j} \leqslant \varphi^q_{m,n,t,j-1}, & \hspace{-3cm} \forall j \in \left\lbrace 2, ..., J \right\rbrace, \label{hq5}\\
& \varphi^q_{m,n,t,j} \in \left\lbrace 0, 1 \right\rbrace, & \hspace{-3cm} \forall j \in \left\lbrace 1, ..., J \right\rbrace.
\end{align}
\end{subequations}

In \eqref{hqequation}, $\delta^q_{m,n,t,j} \in \mathbb{R}$ and $\varphi^q_{m,n,t,j} \in \left\lbrace 0, 1 \right\rbrace$ \vspace{0.1cm} are additional auxiliary continuous and binary variables. The equations \eqref{hq1} and \eqref{hq2} define the linear interpolation curve between the breakpoints via the continuous variable $\delta^q_{m,n,t,j}$ which is constrained to lie between 0 and 1 in equation \eqref{hq3}. Constraints \eqref{hq4} and \eqref{hq5} enforce $\varphi^q_{m,n,t,j}$ equal to 1 or 0, for the intervals on the left or on the right of the current one, respectively. In addition, these equations require that $\delta^q_{m,n,t,j}$ equal to 1 for the intervals on the left, ensuing the proper incremental summation in \eqref{hq1} and \eqref{hq2}. The piecewise linear approximation $h^{pr}_{m,t}$ of square pressures $pr_{m,t}^2$ follows immediately. Given those definitions, the Weymouth equation can be approximated via 
\begin{equation}
h^{q}_{m,n,t} = K_{m,u}^2 \left( h^{pr}_{m,t} - h^{pr}_{n,t} \right) \enspace \forall \left( m,n \right) \in \mathcal{Z}, \; t \in \mathcal{T},
\end{equation}

\noindent which is a linear expression.

\section{Numerical Experiments}\label{sect:NumericalExperiments}





In this section, we summarize the results of numerical experiments performed on a case study inspired by the Belgian power and gas systems. In particular, we solve the proposed neural-network-constrained coordinated power and gas dispatch \eqref{GasElecModel} with the proposed reformulation of activation function in two different settings. The first considers \textit{unidirectional} gas flows (by adding non-negative restrictions $q_{m,n,t} \geqslant 0$, $q^\text{in}_{m,n,t} \geqslant 0$ and $q^\text{out}_{m,n,t} \geqslant 0$), while the second considers \textit{bidirectional} gas flows. The main difference between the two settings stem from the relation captured by the neural network (i.e., unidirectional or bidirectional gas flows). We compare the unidirectional formulation with the existing convexication techniques\footnote{Recall that, unlike our proposed neural-network-constrained framework, it is not straightforward to consider bidirectionality given the state-of-the-art convexification techniques. Therefore, the comparison with respect to those techniques is performed in the unidirectional framework.}. Next, we show the superiority of the proposed neural-network-constrained framework for considering bidirectional gas flow dynamics. All models are either SOC programs or MILPs and are implemented in Julia programming language v1.7.2, using the modeling language JuMP v1.0.0, and the Gurobi solver v9.5.1, on a standard computer clocking at 2.5 GHz, with 16 GB of RAM memory.

The remainder of this section is structured as follows. Section \ref{subsect:CaseStudy} introduces the Belgian power and gas systems, which are used as our case study. Section \ref{subsect:MachineLearning} presents the procedure for training the machine learning regression models. The analysis of the performance of our proposed approach compared to state-of-the-art convexification techniques is derived in Section \ref{subsect:Unidirectional}, given unidirectional gas flow dynamics. The assessment of the proposed reformulation of activation function is discussed in Sections \ref{subsect:LeakyReLU}. Finally, we explore the bidirectional gas flow dynamics in Section \ref{subsect:Bidirectionality}.

\subsection{Case Study: The Belgian Power and Gas Systems}\label{subsect:CaseStudy}

\begin{figure}[]
\subfigure[Belgian electrical grid topology (plain lines are 380 kV and dashed lines are 220 kV).]{\resizebox {0.48\columnwidth} {!} {
\centering
\includegraphics[scale=0.4]{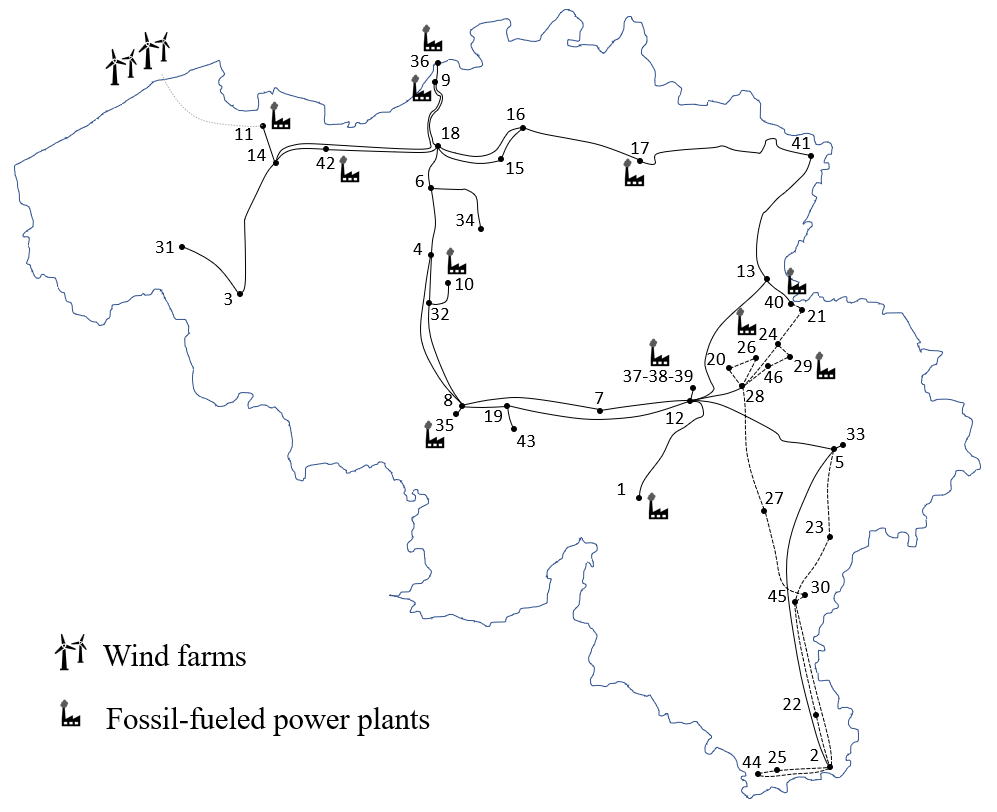}
\label{BelgianPowerTopology}}}
\;
\subfigure[Belgian gas grid topology.]{\resizebox {0.48\columnwidth} {!} {
\centering
\includegraphics[scale=0.4]{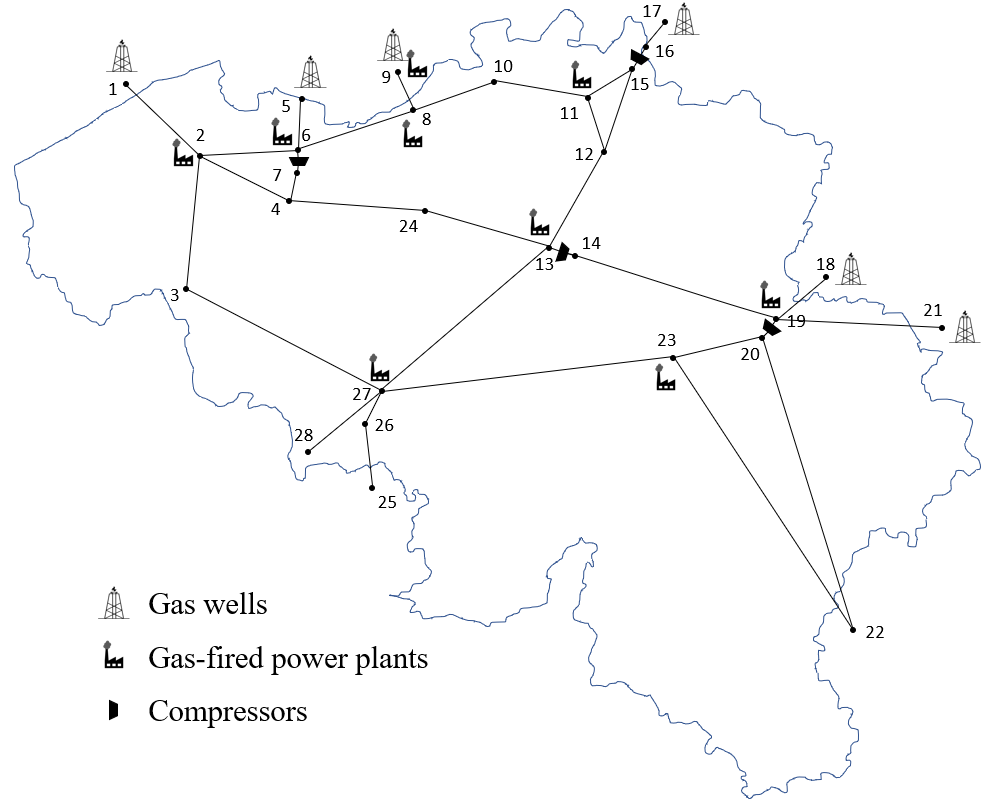}
\label{BelgianGasTopology}}}
\caption{The real-life Belgian power and gas systems and related information. }
\label{fig:inputdata}
\end{figure}

Our case study is based on the Belgian power and gas systems. We consider the high voltage power transmission system, composed of 46 nodes that are connected via 69 transmission lines, as depicted in Fig. \ref{BelgianPowerTopology}. The electrical power demand is supplied by conventional power generating units, NGFPPs, and wind farms, whose aggregate power generating capacities are respectively 6,000 MW, 3,329 MW, and 3,794 MW. The NGFPPs couple the Belgian power and natural gas systems. The latter is a 28-node high pressure gas transmission system, composed of 33 pipelines and 4 compressors, as depicted in Fig. \ref{BelgianGasTopology}. It connects the gas suppliers (with an aggregate capacity of 11.2 MNm$^3$) with the natural gas demand centers. All the technical and economical parameters pertaining to both Belgian power and natural gas transmission systems are reported in our online companion \cite{OnlineAppendix}.


\subsection{Machine Learning Models}\label{subsect:MachineLearning}

The first step for embedding neural networks within the coordinated power and gas dispatch problem is to appropriately select the optimal weight matrices $W$ and $B$ which minimize the value of loss function. \color{black}To do so, we generate $N \geqslant 10^6$ operating data points from the Weymouth surface in Fig. \ref{Fig:3DWeymouth}. \color{black}Next, we split the whole dataset into three subsets, i.e., 60\% of data are used for training (to optimize the model parameters), 20\% are used for validation (for stopping the iterative training procedure at the optimal time, before overfitting issues occur), and the remaining 20\% of data are used as a test set (to compute unbiased performance metrics). 

We repeat this procedure for two different network architectures capturing unidirectional gas flows, i.e., one composed of 2 layers and 5 neurons per layer, and one composed of one layer of 15 neurons. These two architectures are trained with both the ReLU and the leaky ReLU activation functions, resulting in four different neural networks. Note that we arbitrarily impose $\alpha = 0.3$ for the leaky ReLU activation function. Finally, we train four additional neural networks, based on the same characteristics, but capturing bidirectional gas flows. These will be used later in Section \ref{subsect:Bidirectionality}, to enhance the proposed neural-network-constrained framework to bidirectional gas flows dynamics.






\color{black}

\subsection{Ex-post Feasibility under Unidirectional Assumption}\label{subsect:Unidirectional}

\begin{figure*}[]\small
\centering
\subfigure[SOC relaxation. \label{SOC1}]{\resizebox {0.65\columnwidth} {!} {
\begin{tikzpicture}
        \begin{axis}[
            enlargelimits=false,
            ylabel = Pipeline $\mathcal{Z}$,
	        xlabel = Time (h),
	        xmin=1, xmax=24,  
            xtick={3,6,...,24},
            ytick={1,9,17,25,33},
            axis on top,    
            colormap= {bw}{gray(0cm)=(1); gray(1cm)=(0)},
            colorbar,
            colorbar style={
                ylabel= Absolute Error [MNm$^3$/h],
            },
            point meta=explicit,
            point meta min=0,
            point meta max=2,
        ]
\addplot [matrix plot*, point meta=explicit, mesh/cols=24, mesh/rows=33] file [x=Time, y=Pipe] {include/ErrorMatrixData_SOCP.dat};
        \end{axis}
    \end{tikzpicture}}}
    \enspace
    \subfigure[SOC relaxation combined with McCormick envelopes. \label{SOC2}]{\resizebox {0.65\columnwidth} {!} {
\begin{tikzpicture}
        \begin{axis}[
            enlargelimits=false,
            ylabel = Pipeline $\mathcal{Z}$,
	        xlabel = Time (h),
	        xmin=1, xmax=24,  
            xtick={3,6,...,24},
            ytick={1,9,17,25,33},
            axis on top,    
            colormap= {bw}{gray(0cm)=(1); gray(1cm)=(0)},
            colorbar,
            colorbar style={
                ylabel= Absolute Error [MNm$^3$/h],
            },
            point meta=explicit,
            point meta min=0,
            point meta max=2,
        ]
\addplot [matrix plot*, point meta=explicit, mesh/cols=24, mesh/rows=33] file [x=Time, y=Pipe] {include/ErrorMatrixData_SOCP_McCormick.dat};
        \end{axis}
    \end{tikzpicture}}}
    \enspace
    \subfigure[SOC relaxation combined with bound tightening of McCormick envelopes. \label{SOC3}]{\resizebox {0.65\columnwidth} {!} {
\begin{tikzpicture}
        \begin{axis}[
            enlargelimits=false,
            ylabel = Pipeline $\mathcal{Z}$,
	        xlabel = Time (h),
	        xmin=1, xmax=24,  
            xtick={3,6,...,24},
            ytick={1,9,17,25,33},
            axis on top,    
            colormap= {bw}{gray(0cm)=(1); gray(1cm)=(0)},
            colorbar,
            colorbar style={
                ylabel= Absolute Error [MNm$^3$/h],
            },
            point meta=explicit,
            point meta min=0,
            point meta max=2,
        ]
\addplot [matrix plot*, point meta=explicit, mesh/cols=24, mesh/rows=33] file [x=Time, y=Pipe] {include/ErrorMatrixData_SOCP_McCormick_BT.dat};
        \end{axis}
    \end{tikzpicture}}}
    \enspace
    \subfigure[Piecewise linear approximation (5 breakpoints). \label{PLA1}]{\resizebox {0.65\columnwidth} {!} {
\begin{tikzpicture}
        \begin{axis}[
            enlargelimits=false,
            ylabel = Pipeline $\mathcal{Z}$,
	        xlabel = Time (h),
	        xmin=1, xmax=24,  
            xtick={3,6,...,24},
            ytick={1,9,17,25,33},
            axis on top,    
            colormap= {bw}{gray(0cm)=(1); gray(1cm)=(0)},
            colorbar,
            colorbar style={
                ylabel= Absolute Error [MNm$^3$/h],
            },
            point meta=explicit,
            point meta min=0,
            point meta max=2,
        ]
\addplot [matrix plot*, point meta=explicit, mesh/cols=24, mesh/rows=33] file [x=Time, y=Pipe] {include/ErrorMatrixData_PWL.dat};
        \end{axis}
    \end{tikzpicture}}}
    \enspace
      \subfigure[Piecewise linear approximation (10 breakpoints). \label{PLA2}]{\resizebox {0.65\columnwidth} {!} {
\begin{tikzpicture}
        \begin{axis}[
            enlargelimits=false,
            ylabel = Pipeline $\mathcal{Z}$,
	        xlabel = Time (h),
	        xmin=1, xmax=24,  
            xtick={3,6,...,24},
            ytick={1,9,17,25,33},
            axis on top,    
            colormap= {bw}{gray(0cm)=(1); gray(1cm)=(0)},
            colorbar,
            colorbar style={
                ylabel= Absolute Error [MNm$^3$/h],
            },
            point meta=explicit,
            point meta min=0,
            point meta max=2,
        ]
\addplot [matrix plot*, point meta=explicit, mesh/cols=24, mesh/rows=33] file [x=Time, y=Pipe] {include/ErrorMatrixData_PWL119.dat};
        \end{axis}
    \end{tikzpicture}}}
    \enspace
    \subfigure[NN-constrained (2x5, ReLU).]{\resizebox {0.65\columnwidth} {!} {
\begin{tikzpicture}
        \begin{axis}[
            enlargelimits=false,
            ylabel = Pipeline $\mathcal{Z}$,
	        xlabel = Time (h),
	        xmin=1, xmax=24,  
            xtick={3,6,...,24},
            ytick={1,9,17,25,33},
            axis on top,    
            colormap= {bw}{gray(0cm)=(1); gray(1cm)=(0)},
            colorbar,
            colorbar style={
                ylabel= Absolute Error [MNm$^3$/h],
            },
            point meta=explicit,
            point meta min=0,
            point meta max=2,
        ]
\addplot [matrix plot*, point meta=explicit, mesh/cols=24, mesh/rows=33] file [x=Time, y=Pipe] {include/ErrorMatrixData_NNCons_RELU2x5.dat};
        \end{axis}
    \end{tikzpicture}}}
    \enspace
    \subfigure[NN-constrained (1x15, ReLU).]{\resizebox {0.65\columnwidth} {!} {
\begin{tikzpicture}
        \begin{axis}[
            enlargelimits=false,
            ylabel = Pipeline $\mathcal{Z}$,
	        xlabel = Time (h),
	        xmin=1, xmax=24,  
            xtick={3,6,...,24},
            ytick={1,9,17,25,33},
            axis on top,    
            colormap= {bw}{gray(0cm)=(1); gray(1cm)=(0)},
            colorbar,
            colorbar style={
                ylabel= Absolute Error [MNm$^3$/h],
            },
            point meta=explicit,
            point meta min=0,
            point meta max=2,
        ]
\addplot [matrix plot*, point meta=explicit, mesh/cols=24, mesh/rows=33] file [x=Time, y=Pipe] {include/ErrorMatrixData_NNcons_RELU1x15.dat};
        \end{axis}
    \end{tikzpicture}}}
    \enspace
        \subfigure[NN-constrained (2x5, leaky ReLU).]{\resizebox {0.65\columnwidth} {!} {
\begin{tikzpicture}
        \begin{axis}[
            enlargelimits=false,
            ylabel = Pipeline $\mathcal{Z}$,
	        xlabel = Time (h),
	        xmin=1, xmax=24,  
            xtick={3,6,...,24},
            ytick={1,9,17,25,33},
            axis on top,    
            colormap= {bw}{gray(0cm)=(1); gray(1cm)=(0)},
            colorbar,
            colorbar style={
                ylabel= Absolute Error [MNm$^3$/h],
            },
            point meta=explicit,
            point meta min=0,
            point meta max=2,
        ]
\addplot [matrix plot*, point meta=explicit, mesh/cols=24, mesh/rows=33] file [x=Time, y=Pipe] {include/ErrorMatrixData_NNCons_Leaky2x5.dat};
        \end{axis}
    \end{tikzpicture}}}
    \enspace
    \subfigure[NN-constrained (1x15, leaky ReLU).]{\resizebox {0.65\columnwidth} {!} {
\begin{tikzpicture}
        \begin{axis}[
            enlargelimits=false,
            ylabel = Pipeline $\mathcal{Z}$,
	        xlabel = Time (h),
	        xmin=1, xmax=24,  
            xtick={3,6,...,24},
            ytick={1,9,17,25,33},
            axis on top,    
            colormap= {bw}{gray(0cm)=(1); gray(1cm)=(0)},
            colorbar,
            colorbar style={
                ylabel= Absolute Error [MNm$^3$/h],
            },
            point meta=explicit,
            point meta min=0,
            point meta max=2,
        ]
\addplot [matrix plot*, point meta=explicit, mesh/cols=24, mesh/rows=33] file [x=Time, y=Pipe] {include/ErrorMatrixData_NNcons15Leaky.dat};
        \end{axis}
    \end{tikzpicture}}}
     \caption{Matrix plot of the relative error between right-hand and left-hand sides of relaxed Weymouth equations for each time step (x-axis) and each pipeline (y-axis).}
\label{fig:RelativeError}
\end{figure*}
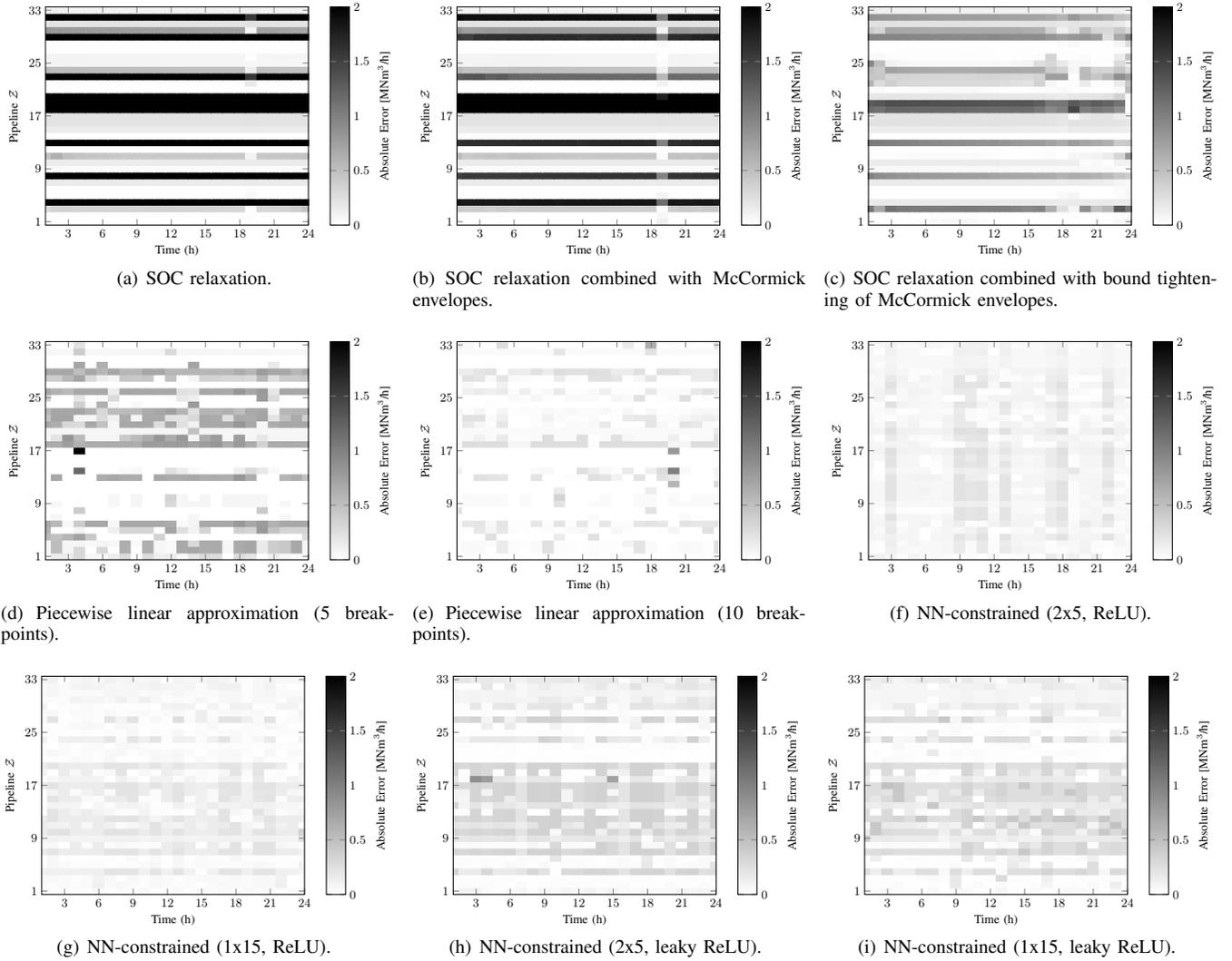

In this section, we aim to evaluate the ex-post feasibility of the obtained neural-network-constrained solutions, compared to existing SOC-based relaxation techniques and the incremental piecewise linear approximation. To do so, we solve the power and gas dispatch with \textit{i)} our proposed neural-network-constrained framework, equipped with the optimal weight matrices $W$ and $B$ of the four architectures discussed in Section \ref{subsect:MachineLearning}, \textit{ii)} the three SOC-based convex relaxation techniques and \textit{iii)} the piecewise linear approximation with two different numbers of breakpoints (i.e., 5 and 10)\footnote{We enforce an optimality gap threshold of 5\% for all the MILP models to solve.}. Next, we retrieve the optimal state variables, i.e., gas flows $q_{m,n,t}$ and pressures $pr_{m,t}$ for all models. We calculate the mismatch with respect to the Weymouth equation \eqref{Weymouth}, as the absolute error between the right-hand side and left-hand side of the Weymouth, i.e., $\Delta_{m,n,t} = \vert q_{m,n,t}^2 - K_{m,n}^2 \left( pr_{m,t}^2 - pr_{n,t}^2 \right) \vert$ for all pipelines $(m,n) \in \mathcal{Z} $and all time periods $t \in \mathcal{T}$, and report the value of absolute errors in the matrix plots of Fig. \ref{fig:RelativeError}. In these figures, darker rectangles highlight those pipelines and time periods with a larger magnitude of error. We also report key performance indicators in Table \ref{Tab:Uni}, such as, the computational time, the Mean Absolute Error (MAE) among all pipelines and all time periods, and the expected total scheduling cost.

\begin{table}[]
\caption{Performance indicators for the neural-network-constrained framework and the benchmark approaches, considering unidirectional gas flows.}
\label{Tab:Uni}
\resizebox {\columnwidth} {!} {
\begin{tabular}{lllll}
\hline
\multicolumn{1}{l|}{Model}                                             & \begin{tabular}{@{}l@{}} Number of \\ binary \\ variables \end{tabular} & \begin{tabular}{@{}l@{}}  Computational \\ time (Gap) [s] \end{tabular}& \begin{tabular}{@{}l@{}} $\text{MAE}_{\text{Opt}}$ \\ ($\text{MAE}_{\text{NN}}$) \\ (MNm$^3$/h)$^2$ \end{tabular} & \begin{tabular}{@{}l@{}} Expected \\ cost [\euro] \end{tabular}\\ \hline

\multicolumn{1}{l|}{SOC}                                &  N.A.                     &  $\leqslant$ 1 s       & 1.54        &  1.41 $\cdot 10^{6}$             \\
\multicolumn{1}{l|}{SOC + McCormick}                    &  N.A.                     &  1.5 s                   &  0.67       &  1.41 $\cdot 10^{6}$             \\
\multicolumn{1}{l|}{SOC + BT$^{1}$} &  N.A.                          & 12 s                    & 0.30       &  1.37 $\cdot 10^{6}$             \\
\multicolumn{1}{l|}{PLA$^{2}$ (5 breakpoints)}                 &  6,528                          &  24 s (0.5 \%)                  &  0.22        &  1.47 $\cdot 10^{6}$             \\
\multicolumn{1}{l|}{PLA (10 breakpoints)}                 &  14,760                         &  60 s (0.7 \%)                  &  0.05     &  1.47 $\cdot 10^{6}$             \\ \hline
\multicolumn{1}{l|}{ReLU, 2x5}                        &  7,920                          &  1622 s (2.7 \%)                & 0.09  (0.045)         & 1.46 $\cdot 10^{6}$              \\
\multicolumn{1}{l|}{ReLU, 1x15}                       &  11,880                          &     905 s (4,4 \%)                & 0.08 (0.038)       &  1.45 $\cdot 10^{6}$             \\
\multicolumn{1}{l|}{Leaky ReLU, 2x5}                  &  7,920                          &    1674 s (4.4\%)                & 0.14 (0.043)     &  1.45 $\cdot 10^{6}$              \\
\multicolumn{1}{l|}{Leaky ReLU, 1x15}                 &    11,880                        &     741 s (4.8 \%)                & 0.13 (0.036)       &  1.44 $\cdot 10^{6}$             \\ \hline
\multicolumn{5}{l}{$^{1}$ BT: Bound tightening \hspace{1cm} $^{2}$: PLA: Piecewise Linear Approximation.}  
\end{tabular}}
\end{table}

One can draw several conclusions from these results. First, the Figs. \ref{SOC1} to \ref{SOC3} confirm that the SOC-based relaxations are not sufficiently tight for a practical implementation (i.e., the MAE achieved by the three SOC-based techniques are the largest). Next, we observe that the proposed neural-network-constrained framework and the incremental piecewise linear approximation achieve comparable values of MAE. In particular, when comparing models involving roughly similar number of binary variables, we conclude that the 2-layer-5-neuron neural networks achieve a lower MAE compared to a piecewise linear approximation with 5 breakpoints, unlike the 1-layer-15-neuron neural networks for which opposite conclusions may be drawn when compared to the piecewise linear approximation with 10 breakpoints. It is worth emphasizing that the matrix plot in Figs. \ref{PLA1} and \ref{PLA2} suggest that the errors from the neural-network-constrained framework are smoother, compared to piecewise linear approximation. The rationale behind this is that the piecewise linear approximation is an interpolation, which may produce large errors locally. In turn, these peak errors may trigger non-trivial operating conditions for system operator. Furthermore, we want to highlight that the neural-network-constrained framework is capable of improving its inherent accuracy by considering real operating measurements in the training phase. In contrast with this interesting feature, the piecewise linear approximation is a model-based solution (i.e., which strongly relies on the Weymouth model) which can not incorporate real operating conditions. Finally, an additional interesting result is that the ReLU activation function achieves lower values of MAE, compared to leaky ReLU activation function, which will be confirmed later in Section \ref{subsect:Bidirectionality} in the case of bidirectional gas flows.

\subsection{Assessing the Proposed Reformulation of Activation Functions}\label{subsect:LeakyReLU}

To assess the benefits of the proposed reformulation of activation function in \eqref{ReforLeakyRelu}, we solve the coordinated power and gas dispatch problem, with the existing reformulation of ReLU defined by \eqref{ReforRelu}. We use the same weight matrices for ReLU as the ones used in Section \ref{subsect:Unidirectional}, and follow the same procedure for reporting the computational time, MAE and expected total scheduling cost in Table \ref{Tab:Spyros}. As our main conclusion, we observe a substantial speed-up, i.e., -37 \% for the 2-layer-5-neuron neural network and -53 \% for the 1-layer-15-neurons one, compared to the previously proposed reformulation. Interestingly, the benefits for utilizing our reformulation increases when the number of neurons involved in the neural network increase. In addition, recall that our proposed reformulation is capable of considering both ReLU and leaky ReLU, though we omit leaky ReLU from this section, for the sake of clarity.


\begin{table}[]
\caption{Performance indicators for the proposed reformulation of ReLU activation function, compared to the existing \cite{SpyrosIEEENN}.}
\label{Tab:Spyros}
\resizebox {\columnwidth} {!} {
\begin{tabular}{llll}
\hline
\multicolumn{1}{l|}{Model}                                             & \begin{tabular}{@{}l@{}}  Computational \\ time (Gap) [s] \end{tabular}& \begin{tabular}{@{}l@{}} $\text{MAE}_{\text{Opt}}$ \\ ($\text{MAE}_{\text{NN}}$) \\ (MNm$^3$/h)$^2$ \end{tabular} & \begin{tabular}{@{}l@{}} Expected \\ cost [\euro] \end{tabular} \\ \hline

\multicolumn{1}{l|}{Model \eqref{ReforLeakyRelu}, 2x5}                                             &  1622 s (2.7 \%)                & 0.086 (0.045)     & 1.46 $\cdot 10^{6}$              \\
\multicolumn{1}{l|}{Model \eqref{ReforLeakyRelu}, 1x15}                                     &     905 s (4,4 \%)                & 0.081 (0.038)       &  1.45 $\cdot 10^{6}$                                               \\ \hline
\multicolumn{1}{l|}{Model \eqref{ReforRelu}, 2x5}                                                &     2602 s (4.7 \%)                & 0.070 (0.045)        &    1.48 $\cdot 10^{6}$           \\
\multicolumn{1}{l|}{Model \eqref{ReforRelu}, 1x15}                                   &     1954 s (4.9 \%)                & 0.075 (0.038)       &   1.45 $\cdot 10^{6}$            \\ \hline
\end{tabular}}
\end{table}

\subsection{Performance under Bidirectionality Assumption}\label{subsect:Bidirectionality}

In this section, we enhance the neural-network-constrained power and gas dispatch with bidirectional gas flow dynamics. To do so, we train the neural network based on bidirectional operating data points, and simply plug the weight matrices in a power and gas dispatch problem where the gas flow state variables (i.e., $q_{m,n,t}$, $q_{m,n,t}^\text{out}$ and $q_{m,n,t}^\text{in}$) can either be positive or negative. We solve the bidirectional power and gas dispatch with the four different architectures of neural networks discussed in Section \ref{subsect:MachineLearning} and report the performance indicators in Table \ref{Tab:Bidir}. Our main observation is that the MAE of the obtained solution is consistent with those obtained with the unidirectional model (i.e., in general, the MAE slightly decreases). However, the expected total scheduling cost is reduced by about 20 \% on average for the neural networks considered in this study. The rationale behind this is that the model with bidirectional gas flow dynamics provides more flexibility to the system operator for scheduling the system, therefore revealing new opportunities to schedule the cheapest units. An additional interesting element is that the MAE of the obtained optimal solution and that obtained by the neural-network transfer function \textit{per se} are different (e.g., the 2-layer-5-neuron neural network based on ReLU achieves a MAE of 0.326 but the obtained scheduling solution attains a MAE equal to 0.08). This observation sheds light on the importance for the regression model to be accurate in the operating zones that are exploited within the optimization problem. 


\begin{table}[]
\caption{Performance indicators for the neural-network-constrained framework, considering bidirectional gas flows.}
\label{Tab:Bidir}
\resizebox {\columnwidth} {!} {
\begin{tabular}{lllll}
\hline
\multicolumn{1}{l|}{Model}                                             & \begin{tabular}{@{}l@{}} Number of \\ binary \\ variables \end{tabular} & \begin{tabular}{@{}l@{}}  Computational \\ time (Gap) [s] \end{tabular}& \begin{tabular}{@{}l@{}} $\text{MAE}_{\text{Opt}}$ \\ ($\text{MAE}_{\text{NN}}$) \\ (MNm$^3$/h)$^2$ \end{tabular} & \begin{tabular}{@{}l@{}} Expected \\ cost [\euro] \end{tabular} \\ \hline

\multicolumn{1}{l|}{ReLU, 2x5}                        &  7,920                          & 1257 s (3.9 \%)                    &  0.08 (0.326)     &  1.13 $\cdot 10^{6}$             \\
\multicolumn{1}{l|}{ReLU, 1x15}                       &   11,880                         &  7200 s (21.3 \%)                   & 0.08 (0.533)      & 1.23 $\cdot 10^{6}$              \\
\multicolumn{1}{l|}{Leaky ReLU, 2x5}                  &  7,920                          &  1558 s (4.0 \%)                   & 0.11 (0.491)        &  1.14 $\cdot 10^{6}$            \\
\multicolumn{1}{l|}{Leaky ReLU, 1x15}                 &  11,880                          &  4317 s (2.8 \%)                   & 0.12 (0.453)     &  1.12 $\cdot 10^{6}$              \\ \hline
\end{tabular}}
\end{table}




\section{Conclusions}\label{sec_concl}






In this paper, to cope with the non-convexity stemming from the modeling of gas flow dynamics within the coordinated power and gas dispatch, we propose a neural-network-constrained optimization method which includes a regression model of the Weymouth equation into the set of constraints. In addition, we introduce a reformulation of the activation function which improves the computational efficiency. Through several numerical experiments, we show that our framework is capable of recovering a tractable MILP, which outperforms the previously inaccurate SOC-based convexification techniques and is comparable to piecewise linear approximation methods. Interestingly, the proposed neural-network-constrained framework is capable of improving accuracy, without relying on analytical assumptions, by considering real operating measurements.

As future research work, we highlight the potential of our proposed neural-network-constrained framework to consider other non-convex models, e.g., the heat system operation which is described by the non-convex Darcy-Weidbach equations, linking pressures to mass flow rates. Another potential extension is linked to machine learning probabilistic regression tools. Following the same idea, neural-network constraints could be leveraged to model non-convex probabilistic constraints. Finally, it is interesting to apply the proposed framework within regulatory-compliant coordination schemes to reveal the benefits of the approach in these different settings.

\bibliographystyle{IEEEtran}
\bibliography{IEEEabrv, full_paper}

%
%

\end{document}